\documentclass[aps,prb,twocolumn, showpacs]{revtex4}

\usepackage{epsfig}
\usepackage{graphicx}

\def\be{\begin{equation}}
\def\ee{\end{equation}}
\def\bea{\begin{eqnarray}}
\def\eea{\end{eqnarray}}

\begin{document}

\title{Mechanism of half-frequency electric dipole spin resonance in double quantum dots:\\
Effect of nonlinear charge dynamics inside the singlet manifold}
\author{Emmanuel I. Rashba}
\affiliation{Department of Physics, Harvard University, Cambridge, Massachusetts 02138, USA}  
\date{\today}
\begin{abstract}
Electron dynamics in quantum dots manifests itself in spin-flip spectra through electric dipole spin resonance (EDSR). Near a neutrality point separating two different singlet charged states of a double quantum dot, charge dynamics inside a $2\times2$ singlet manifold can be described by a $1/2$-pseudospin. In this region, charge dynamics is highly nonlinear and strongly influenced by flopping its soft pseudospin mode. As a result, the responses to external driving include first and second harmonics of the driving frequency and their Raman satellites shifted by the pseudospin frequency. In EDSR spectra of a spin-orbit couplet doublet dot, they manifest themselves as charge satellites of spin-flip transitions. The theory describes gross features of the anomalous half-frequency EDSR in spin blockade spectra [Laird et al., Semicond. Sci. Techol. {\bf 24}, 064004 (2009)]. 
\end{abstract}

\pacs{72.25.-b,73.21.La,75.70.Tj,85.75.-d}

\maketitle

\narrowtext

Operation of electron spins in quantum dots is critical for spintronics\cite{Zutic} and especially for quantum computing.\cite{LDV,KouMar98,Awsch02} Electron spins can be operated by electron spin resonance (ESR) driven by a time-dependent magnetic field $\tilde{\mbox{\boldmath$B$}}(t)$\cite{Koppens06} or by electric dipole spin resonance (EDSR) driven by a time-dependent electric field $\tilde{\mbox{\boldmath$E$}}(t)$.\cite{R60,EDSR,Golovach06,Laird07,Nowack07,Tarucha08} As applied to nanostructures, EDSR is easier to operate because it only requires voltages applied to gates rather than striplines and is usually more efficient than ESR. While the physical mechanisms of EDSR differ depending on the coupling of the electron spin ${\bf s}=\frac{1}{2}\mbox{\boldmath$\sigma$}$ to the electron momentum $\hat{\bf k}$ or to its coordinate $\bf r$ through the spatially dependent Zeeman energy in inhomogeneous magnetic field ${\bf B}({\bf r})$ (external, exchange, or hyperfine),\cite{EDSR,KMDGLA03,RashbaJSup2005,Tokura06} the frequencies $\omega$ of both ESR and EDSR are equal to the Zeeman energy $\omega_B=|g|\mu_BB/\hbar$ due to the energy conservation.

However, more recently Laird et al.\cite{Laird2009} reported intensive EDSR in GaAs double quantum dots (DQD) at the {\it half-frequency} $\omega=\frac{1}{2}\omega_B$ observed side by side with the EDSR at the basic frequency $\omega=\omega_B$;\cite{Laird2009} it was detected by breaking Pauli blockade. They emphasized that a ``half-frequency response is as far as we know unprecedented in spin resonance". Indeed, such a response requires strong nonlinearity that is unknown for spin resonance in its previous realizations. Also, such a behavior has no similarity in nonlinear spectroscopy where absorption of the first and second harmonics is alternatively forbidden in centrosymmetric systems by parity conservation, a requirement reasonably well fulfilled in symmetric single quantum dots. 

Analysis of experimental data\cite{Laird2009} implies a mechanism of this unique behavior. Most important, the half-frequency resonance was observed only in the spin blockade region close to the degeneracy point of the singlet (1,1) and (0,2) charge states of DQD.\cite{CD,vanderWiel2003} This observation suggests that the {\it mechanism of nonlinearity was electrical} and stemmed from the fact that the DQD was {\it close to the resonance between two singlet configurations} where the polarizability of the system was {\it high}\cite{Frey2011} and {\it strongly nonlinear}. Also, the resonance was only detected in a device including a micromagnet. This observation indicates that the coupling of charge and spin dynamics  underlying the EDSR was through $\bf r$-dependent Zeeman energy. Indeed, the topography of stray field [Fig.~5(b) of Ref.~\onlinecite{Laird2009}] shows that ${\bf B}({\bf r})$ changed considerably at the DQD-size scale. However, this specific mechanism of spin-orbit coupling is not essential for what follows, and in InAs where intrinsic spin-orbit coupling is two orders of magnitude stronger than in GaAs, similar effects are expected even in the absence of micromagnet.

In what follows, I calculate single- and double-frequency electrical responses of a DQD in terms of electric dipole moments ${\bf r}_\omega(t)$ and ${\bf r}_{2\omega}(t)$ induced by a harmonic $\omega$-frequency perturbation. The result is specific for DQD's near the intersection of (1,1) and (0,2) energy levels where the system is close to degeneracy; see Fig.~1. Away from this intersection, both dots react to the perturbation independently, and 
\be
{\bf r}(t)=-e\tilde{\mbox{\boldmath$E$}}(t)/m\omega_0^2, 
\label{eq0}
\ee
where $\omega_0$ and $m$ are the confinement frequency and electron mass, respectively.\cite{Golovach06,Laird07,Rashba08} Because of the large $\omega_0$, the displacement of electron density is small compared with the dot size $a\approx\sqrt{\hbar/m\omega_0}$, nonlinear effects can be disregarded, and higher harmonics do not develop.

The specific form of the charge stability diagram of a DQD in a wide parameter range depends on the shape of the dot, the potentials on the gates, and the electron-electron interaction inside the dot. However, as applied to nonlinear responses that are the focus of the present Rapid Communication, only narrow regions near the charge-balance lines are of actual importance. In such narrow regions, all detailed parameters of the DQD can be absorbed into a few parameters of a model Hamiltonian. To keep connection to the data of Ref.~\onlinecite{Laird2009}, I concentrate on the vicinity of the charge balance line separating the stability regions of (0,2) and (1,1) singlet states that are displayed schematically in Fig.~1. In this figure, the charge balance line is shown as a blue (heavy)  line, and charge dynamics is considered in the direction perpendicular to this line. In the vicinity of the line, and not too close to its ends, the Hilbert space of the system is spanned by two charge states, (0,2) and (1,1).

In the (0,2) and (1,1) basis, the Hamiltonian of a two-electron DQD is
\be
H=-[\epsilon+h(t)]\tau_z+v\tau_x,
\label{eq1}
\ee
where $\mbox{\boldmath$\tau$}$ is a vector of Pauli matrices acting in the singlet (0,2) - (1,1) subspace; the corresponding degree of freedom will be termed a pseudospin in what follows. 
Here $\epsilon$ is the static detuning equal to a half-difference of the energies of (1,1) and (0,2) states, $v$ is the tunneling matrix element, and
\be
h(t)=h\cos\omega t
\label{eq2}
\ee
is the time-dependent detuning. Because all dynamics proceeds inside the singlet subspace, the real spin is suppressed.

Static part $H_0$ of the Hamiltonian $H$ can be diagonalized by an unitary transformation
\be
U=\alpha\tau_z+\beta\tau_x,\,\alpha=\sqrt{(\Delta-\epsilon)/2\Delta},\,\beta=\sqrt{(\Delta+\epsilon)/2\Delta}
\label{eq3}
\ee
with the eigenvalues $\epsilon_\pm=\pm\Delta,\,\,\Delta=\sqrt{\epsilon^2+v^2}$. Therefore, $2\Delta$ is the excitation energy of the pseudospin degree of freedom. The sum $J=\Delta+\epsilon$ is usually considered as the exchange energy separating $T_0$ triplet and the lower singlet $S$ state.\cite{Petta2005}

The transformed total Hamiltonian $H_U=UHU^{-1}$ is
\be
H_U=\Delta\tau_z+(\epsilon h(t)/\Delta)\tau_0+(vh(t)/\Delta)\tau_x\,,
\label{eq4}
\ee
where $\tau_0$ is a unit matrix in \mbox{\boldmath$\tau$} space, and the dipole moment $r(t)$ of DQD can be expressed in terms of the solution $b(t)=(b_1(t),b_2(t))^T$ of a Schroedinger equation
\be
i\partial_tb(t)=H_Ub(t);
\label{eq5}
\ee
here and below $\hbar=1$. This dipole moment $r(t)=r^0(t)+r^S(t)$ consists of two terms, of which the first
\bea
r^0(t)&=&\{\alpha^2\vert b_1(t)\vert^2+\beta^2\vert b_2(t)\vert^2\nonumber\\
&+&\alpha\beta[b_1(t)b_2^*(t)+b_2(t)b_1^*(t)]\}Z
\label{eq6}
\eea
is large and scales with the interdot separation $Z$. The second term 
\bea
r^S(t)&=&\{2\alpha\beta(\vert b_1\vert^2-\vert b_2\vert^2)\nonumber\\
&-&(\alpha^2-\beta^2)[b_1(t)b_2^*(t)+b_2(t)b_1^*(t)]\}Z_S
\label{eq7}
\eea
scales with $Z_S=\langle\psi_{02}\vert z_1+z_2\vert\psi_{11}\rangle$, where $\psi_{02}$ and $\psi_{11}$ are wave functions of the singlet (0,2) and (1,1) states, respectively. This term is of the order of $SZ$, where $S\ll1$ is the overlap integral of the single-electron wave functions $\psi_L$ and $\psi_R$ of the left and right dots. The $z$ direction is chosen along the axis of DQD.

Equation (\ref{eq5}) cannot be solved analytically, and experimental data suggest a regime in which the frequency of the perturbation $\omega$ is large compared the energy scales of the static Hamiltonian, $\omega\gg v,\epsilon$. Indeed, this is a necessary condition for the energies of spin-blocking states $(T_+,T_-)$ be separated from unblocked singlet $S$ and triplet $T_0$ states by energies close to $\omega_B$. Therefore, in what follows I perform an expansion in $1/\omega$ and consider $\Delta/\omega, h/\omega\ll1$ as small parameters, while $h$ and $\Delta$ may be of the same order of magnitude. 

To this end, the diagonal $\pm \Delta$ terms in Eq.~(\ref{eq4}) can be eliminated by a substitution
\be
b_1(t)=c_1(t)e^{-i\Delta t},\,\,b_2(t)=c_2(t)e^{i\Delta t}.
\label{eq8}
\ee
Then nondiagonal terms of $H_U$ acquire factors $\exp(\pm 2i\Delta t)$, and all terms of the Hamiltonian become proportional to $h$. This transformation allows to take into account, in a consistent way, the slowly oscillating factors that become turn responsible for the Raman satellites. 

Next, both functions $c_m(t)$, $m=\{1,2\}$, are expanded in Fourier series
\be
c_m(t)=c_{m0}+\sum_{n>0}(c_{mn}^+e^{in\omega t}+c_{mn}^-e^{-in\omega t}),
\label{eq9}
\ee
where $c_{mn}$ depend on $t$ but slowly compared with $e^{\pm i\omega t}$. Because the usual temperature of the thermostat $\sim$ 100 mK $\approx $ 10$\mu$eV is of the same order of magnitude as $\Delta$, the state $(c_{10},c_{20})$ is a mixed one even when $h=0$. It can be parameterized as $c_{10}=\cos\varphi, c_{20}=\sin\varphi e^{i\phi}$, and this choice is general enough to provide results in terms of the density matrix approach. While the driving term $h(t)$ results in a slow $t$-dependence of $(c_{10},c_{20})$, it can be disregarded in the calculations performed below.

For calculating the response $r(t)$ including the single- and double-frequency terms, dynamic equations for first two coefficients $c_{mn}(t)$, with $n=1$ and $n=2$, were solved by expanding in $1/\omega$. The solutions for $c_{mn}^\pm$ include time-independent terms and terms oscillating as $e^{\pm 2i\Delta t}$. The appearance of the Raman satellites is seen most clearly from the explicit form of the Fourier components $b_{11}^\pm=c_{11}^\pm e^{-i\Delta t}$ and $b_{21}^\pm=c_{21}^\pm e^{i\Delta t}$
\bea
b_{11}^\pm&=&\mp\frac{\epsilon}{2\Delta}\frac{h}{\omega}e^{-i\Delta t}\pm\frac{v}{2\Delta}\frac{h}{\omega}e^{i\Delta t}e^{i\phi}\sin\varphi,\nonumber\\
b_{21}^\pm&=&\pm\frac{v}{2\Delta}\frac{h}{\omega}e^{-i\Delta t}\pm\frac{\epsilon}{2\Delta}\frac{h}{\omega}e^{i\Delta t}e^{i\phi}\sin\varphi,
\label{Raman}
\eea
that include both $e^{\pm i\Delta t}$ harmonics. Next, the Fourier components, $b^\pm_{12}$ and $b^\pm_{22}$, can be expressed in a similar way in terms of $b^\pm_{11}$ and $b^\pm_{21}$ with corresponding time dependent coefficients. Expressions for the $\omega$- and $2\omega$-responses of Eqs.~(\ref{eq6}) and (\ref{eq7}) are more cumbersome because they include biquadratic forms of the amplitudes $b_1(t)$ and $b_2(t)$. 

Before discussing specific results for  these responses, it is instructive to take into account that $r^0(t)$ satisfies, due to the equations of motion of Eq.~(\ref{eq5}), an exact identity
\be
i\partial_t r^0(t)=v(b_1b_2^*-b_1^*b_2)Z.
\label{eq10}
\ee
The derivative $\partial_t r^0(t)$ is proportional to $\omega$ both for the first and second harmonic responses. Therefore, the $r^0(t)$ term in $r(t)$, while scaled with a large length $Z$, includes an additional $\omega$ factor in the denominator and therefore decreases with $\omega$ faster than $r^S(t)$.

The $\omega$-responses calculated using Eqs.~(\ref{eq6}) and (\ref{eq7}), in the leading order in $1/\omega$, are
\be
r^0_\omega(t)=\frac{2hv}{\Delta\omega^2}[v\cos2\varphi+\epsilon\sin2\varphi\cos(\phi+2\Delta t)]Z\cos\omega t ,
\label{eq11}
\ee
\be
r^S_\omega(t)=-\frac{2h}{\omega}\sin2\varphi\sin(\phi+2\Delta t)Z_S\sin\omega t .
\label{eq12}
\ee
Therefore, the linear response includes oscillations at the driving frequency $\omega$ and two satellites at frequencies $\omega\pm2\Delta$. The satellites originate from slowly oscillating factors $e^{i\Delta t}$ in the Fourier amplitudes of the same type as in Eq.~(\ref{Raman}). To evaluate the relative magnitude of both contribution, one can estimate $v\sim S\varepsilon_0$, where $\varepsilon_0\sim$ 1 meV is the intradot electron energy. Then $r^S_\omega/r^0_\omega\sim\omega_B/\varepsilon_0$. With $\omega_B\sim$ 2.5 $\mu$eV for GaAs at $B\sim100$ mT, $r^0_\omega$ appears as a dominating contribution despite its faster decrease with $\omega$.

Double frequency responses calculated in a similar way are
\be
r^0_{2\omega}(t)=-\frac{h^2v}{2\omega^3}\sin2\varphi\sin(\phi+2\Delta t)Z\sin2\omega t,
\label{eq13}
\ee
\be
r^S_{2\omega}(t)=\frac{h^2}{\Delta\omega^2}[v\cos2\varphi+\epsilon\sin2\varphi\cos(\phi+2\Delta t)]Z_S\cos2\omega t.
\label{eq14}
\ee
For the reasons explained above, the $r^0_{2\omega}$ term is expected to dominate, and it does not include the unshifted (central) second-harmonic contribution but only a doublet of Raman terms with the frequencies $2(\omega\pm \Delta)$. Meanwhile, $r^S_{2\omega}$ includes a central $2\omega$ line and its $2(\omega\pm \Delta)$ satellites. 

Notably, $2\Delta$ energy shifts of the oscillation frequencies are universal for all satellite bands, both in the linear and quadratic responses. Because $2\Delta$ is the energy separation between  $\epsilon_\pm$ sublevels of the static Hamiltonian $H_0$, the origin of these shifts can be easily understood. Physically, they come from the flopping between two pseudospin levels driven by the high-frequency perturbation $h(t)$. Technically, they originate in the biquadratic expressions of Eq.~(\ref{eq6}) and (\ref{eq7}) from multiplication of $e^{\pm i\Delta t}$ factors in the amplitudes $b(t)$.  

A comparison of Eqs.~(\ref{eq11}) and (\ref{eq12}) and Eqs.~(\ref{eq13}) and (\ref{eq14}) shows that the ratio $r_{2\omega}/r_{\omega}$ is of the order of $h/\omega$, i.e., it is small in the parameter $h/\omega$  used above for deriving all equations. However, because $h$ appears only in prefactors rather than in the functional dependencies, it is probable that all qualitative conclusions are valid until $h\alt\omega$. While the magnitude of the third harmonic has not been calculated explicitly, its magnitude (and therefore the intensity of the $\omega=\frac{1}{3}\omega_B$ resonance) is expected to include an additional power of $h/\omega$.  

Coexistence in a geometrically symmetrical DQD of the first and second harmonics stems from the fact that electronic density near the charge balance point of (0,2) and (1,1) configurations is asymmetrical. While the elastic term in the nonlinear response of Eq.~(\ref{eq14}) may be, at least partially, attributed to the static asymmetry of electron density distribution, the Raman shifted terms originate completely from charge dynamics in the pseudospin space. Dynamical asymmetry manifests itself explicitly in strong parity violation of $2\Delta$ satellites that are ubiquitous and appear both in the $\omega$ and $\omega^2$ responses.

All dipole moments in Eqs.~(\ref{eq11}) and (\ref{eq14}) vanish after averaging over $\varphi$. This means that their amplitudes and phases strongly depend on the quantum state $(c_{10},c_{20})$. Whenever the dynamics driven by the field $h(t)$ is faster than the change of this state controlled by its coupling to the bath, both linear and nonlinear polarizabilities remain unchanged. On a longer time scale the polarizabilities fluctuate and may be estimated by their mean square values. 

Now we are in a position to make connection to the data of Ref.~\onlinecite{Laird2009}. According to the above assumptions, oscillations of electronic density parameterized above in terms of the dipole moments $r(t)$ drive EDSR through the ${\bf B}({\bf r})$ mechanism and in this way affect the efficiency of the Pauli blockade. In Fig.~6 of Ref.~\onlinecite{Laird2009} both the linear and quadratic responses are split, at least into two components (visibility of the components is reduced by fluctuations of the nuclear bath). At $B\approx100$ mT, where this splitting is best seen, it is $\sim$ 10\% of the Zeeman energy. Because the splitting does not increase with the external magnetic field $B$, it can not be related to the difference in the $g$-factors of two dots. Therefore, it is natural to attribute it to the satellite structure unveiled above, which physically corresponds to two charge (pseudospin) Raman-type satellites accompanying each spin-flip line. Estimating the splitting as $4\Delta$ in the spirit of Eqs.~(\ref{eq11})-(\ref{eq14}), we arrive at $\Delta\sim0.1 \mu$eV which corresponds to tiny tunnel matrix element $v$ and detuning $\epsilon$.  The visibility of both resonances, especially of the half-frequency $\omega=\frac{1}{2}\omega_B$ resonance, decreases with $B$, in a qualitative agreement with the decrease of the dipole-moment matrix elements of Eqs.~(\ref{eq11})-(\ref{eq14}). 

While the gross features of the exotic EDSR observed in Ref.~\onlinecite{Laird2009} are in a reasonable agreement with the above theory, the real situation is more complicated. Indeed, while the $\omega=\omega_B$ resonance manifests itself as a peak in the current, the $\omega=\frac{1}{2}\omega_B$ resonance is seen as a dip in the background current leaking through the Pauli blocked DQD. Most probably, this indicates that the half-frequency feature appears as a Fano resonance,\cite{NSP} and therefore different mechanisms of breaking the spin blockade (that is weak in the vicinity of the charge degeneracy point) should be taken into account. Moreover, signs of both resonances change depending on the exact gate configuration,\cite{Laird2009} which is not surprising having in mind the tiny magnitude of $\Delta$ estimated above. Detailed patterns of the Pauli blocked currents near the charge balance point are outside the scope of the current paper.

The mechanism of EDSR described above allows for (i) a new spectral region of the operation of singlet-triplet DQD qubits and (ii) a strong and controlled dynamic coupling between them.

In conclusion, near the charge neutrality point the polarizability of a double quantum dot is large, highly nonlinear, and strongly influenced by dynamics of its soft pseudospin mode. A high-frequency detuning signal $h(t)=h\cos\omega t$ drives dynamics inside the $2\times2$ singlet subspace producing flops of the pseudospin mode between its two eigenstates separated by energy $2\Delta\ll\omega$. This charge dynamics results in oscillations of electron density at the frequencies $\omega$ and $2\omega$ and at the Raman shifted frequencies $\omega\pm 2\Delta$ and $2\omega\pm2\Delta$. Pseudospin flops also act as an efficient mechanism of parity violation. Through one of the mechanisms of spin-orbit coupling, the density oscillations drive EDSR, and shifted frequencies manifest themselves as Raman-type satellites revealing the pseudospin charge dynamics in spin-flip spectra.  

I am grateful to A. Higginbotham, F. Kuemmeth, and C. M. Marcus for inspiring discussions and acknowledge funding from the Intelligence Advanced Research Project Activity (IARPA), through the Army Research Office, and by NSF under Grant No. DMR-0908070.

\vspace{5mm}

{\bf Caption to Fig.~1}

Fig.~1. (Color online) Charge-balance diagram in the vicinity of the (1,1) - (0,2) charge-balance line shown in blue (heavy solid). Both time-independent $\epsilon$ and time-dependent $h(t)$ detunings are applied along the horizontal dashed line marked as $\epsilon$. Along this line, charge dynamics within the $2\times2$ (1,1)-(0,2) singlet subspace is described in the framework of a pseudospin-$1/2$ formalism.


\begin{thebibliography}{99}
\bibitem{CD} In these states, two-electron density is, respectively, distributed symmetrically between both dots or concentrated on the right dot.
\end{thebibliography}
\end{document}